\documentclass{hch}
\usepackage{graphicx}

\newcommand{\dgam}{\dot{\gamma}}             
\newcommand{\dGam}{\dot{\Gamma}}

\begin{document}

\title{Mechanical aging and non-linear rheology 
of concentrated colloidal suspensions: experimental facts and simple models.}

\author{Armand Ajdari}
\thanks{Great thanks 
to Lyd\'eric Bocquet, Michael Cates, Caroline Derec, David Head, Fran\c cois Lequeux and 
Guillemette Picard, 
for past and ongoing collaboration on related topics.}
\address{Physico-Chimie Th\'eorique, \\
UMR CNRS-ESPCI 7083, \\
  10 rue Vauquelin, 75005 Paris, France}
\runningtitle{Mechanical aging and non-linear rheology}
\maketitle
\begin{abstract}\hskip 0.15in 
Many colloidal systems display very non-Newtonian and solid-like behaviour when concentrated,
a striking feature being the apparition of a yield stress. After recalling 
some basics about the interactions between colloidal particles,
I present a few experimental facts commonly observed in these systems:
aging and non-linear rheology. A simple
phenomenological model is then introduced, in which the local state of
the system is described by a single scalar parameter, the fluidity.
I proceed with comments on heterogeneous flows in some of these systems.
These notes are not intended to be a comprehensive 
review, and the reader is directed to the references for further reading.
\\
%
\end{abstract}
\vspace*{-0.5cm} 

\date{Notes for lectures in les Houches 2002}
\maketitle

\section{Introduction}
\subsection{Colloidal glasses ?}

In the last years many works \cite{sol1,liu1,clo1,der1,wei1,ber1}
have suggested possible cross-interests between
(i) research on the glass transition (statistical mechanics, structural glasses,
solid polymers), (ii) studies on colloidal ``soft'' systems .
Such cross-interests are based on the two following hopes.

First, colloidal systems could be ideal systems to understand fundamental
aspects of the glass transition: 
they are
{\em tunable} ``equivalents'' of atomic or molecular systems
with controllable interactions,
that also sometimes freeze into amorphous disordered phases with solid-like behaviour 
(``glasses'').
The spatial scale is advantageously roughly a thousand times larger than in atomic systems, 
which could permit easier experimental 
investigation. In addition, playing with the viscosity of the solvents,
one can hope to tune the temporal scales, allowing an easier span of the
huge range of accessible time scales required for the study of the glass transition.

Second, many common colloidal systems (gels, foams, creams, emulsions)
are actually used
in this soft solid form that may bear similarities with glasses.
Mayonnaise for example is a disordered arrangement of oil drops
that behave as an amorphous elastic solid  under a weak constraint,
and flows plastically under larger stress.
Some of the mechanical features of these 
amorphous colloidal systems may be better apprehended 
if they are considered as glasses.

\subsection{Model and real colloids: interactions}

A first step toward the understanding of the behaviour of a colloidal system 
is the examination of the interactions between its constituents.
Without aiming at an exhaustive review, I describe below 
a set of commonly encountered attractions and repulsions 
between colloidal particles embedded in a fluid solvent (thorough reviews
are provided by textbooks such as \cite{rus1,isr1} ).
The term ``colloid'' is often coined to particles of
size smaller than a few microns dispersed in a solvent, 
so that gravity has a 
limited effect on a {\em single} particle (the exact limit depending
on the experiment or application considered). This allows the dispersion of the
particles by simple thermal agitation.

\subsubsection{Interactions between model colloids}

\begin{figure}[t]
\vspace{.3cm}
\centering
\includegraphics[width=10cm]{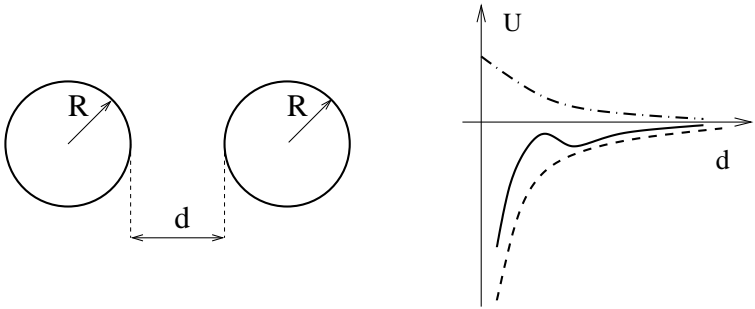}
\caption{{\em Left :} Two model spherical colloids. {\em Right :}
Typical pair potentials in the DLVO theory: the van der Waals attraction 
(dashed line) and the electrostatic repulsion (dot-dashed line) result in an overall
interaction potential (solid line) with a secondary minimum at a a finite distance $d$. 
}
\vspace{-.3cm}
\label{fig1}
\end{figure}

{\em Attractions and Repulsions -} I consider here 
two identical colloidal spheres of radius $R$ in a solvent, 
separated by a distance $d$ (Figure \ref{fig1}). They induce on each other
both attractions (A) and repulsions (R):

A1- The van der Waals attraction, 
due to the difference in dielectric properties of the solvent and the particles. 
This leads to a long-range attractive potential $U_{\rm vdw}(d)$, that, in a simple picture 
neglecting retarded interactions, behaves
as $\sim - H d/R$ for $d \ll R$ and $\sim - H (d/R)^6$ for $d \gg R$.
The Hamaker constant $H$ is roughly proportional to the square of the difference in dielectric constant
(or in optical index) and of order $k_BT$. 
By fine-tuning solvent mixtures it is sometimes possible
to diminish the contrast and significantly reduce this attraction.

A2 -  Depletion forces,
that can for example be induced by dissolving in the solvent
polymers that do not adsorb on the surface.
The dissolved polymer coils behave as impenetrable objects of typical size
say $b$, which are sterically excluded (depleted) from a corona around the colloidal particles
of a thickness of order $b$. Bringing particles together induces overlap of these
depletion coronas, and thus increases the volume
accessible to the polymers and consequently the configurational entropy of the system. This
induces {\em a short-range ``depletion'' attraction} 
of entropic origin. Its range is of order $b$ and its amplitude controlled by
the polymer concentration.

R1 - Core steric repulsion: the particles cannot interpenetrate which results 
in a very strong and very short-ranged repulsion (a few Angstroms).

R2 - Screened electrostatic repulsion:
similarly charged particles dispersed in an electrolyte solvent (e.g. water)
repel each other.
Electrostatic effects are screened over the Debye length $\lambda_D$ 
(a few $nm$ to a few tens of $nm$ typically),
due to the presence of dissociated counter-ions and salt. 
The resulting repulsion potential $U_{\rm el}(r)$ decreases
as $\sim \exp(- d/\lambda_D)$ for large separations. 
Increasing the ionic strength of the solution (e.g. by adding salt)
induces a drop of both the amplitude and range $\lambda_D$ of this repulsion.

R3- (Steric) protection by polymer layers.
 A classical way to avoid aggregation of the particles 
is to sterically ``protect'' the colloids with a polymer layer of thickness $\delta$ (grafted or adsorbed).
Such layers swollen by the solvent repel each other for entropic reasons
and thus provide a rather strong wall potential at $d\simeq 2\delta$. In addition, 
the dielectric properties
of these layers of low polymer concentration
are close to that of the solvent, so that they do not induce unfavorable additional van der Waals 
attractions.  \\

{\em Resulting pair potentials -} In general many interactions act concurrently. 
The classical DLVO theory
describes the behaviour of simple charged colloidal spheres in a salty solution 
under the action of van der Waals attraction and electrostatic repulsions
(A1, R1 and R2). 
The attraction always dominate
at very long and very short separations, and the electrostatic repulsion if strong enough can 
induce  a ``secondary'' potential minimum (see Figure \ref{fig1}). The overall behaviour 
of a suspension of such particles then depends on the depth of this minimum 
and on the amplitude of the barrier that separates it from the ``primary'' minimum (which
corresponds to quasi-contact).
If particles reach this quasi-contact state, 
very short range interactions and chemistry often take over
that result in an irreversible binding.

If the attraction wins, for example upon addition of salt to the system,
the particles tend to aggregate into big lumps that sediment or cream (gravity
is far from negligible for large aggregates),
leading to the ``destabilization'' of the suspension, an effect often fought against
(for example in paints, inks, etc ..).
Protection techniques have often been used to strengthen repulsion (R3)
and prevent actual contact.

\begin{figure}[t]
\vspace{.3cm}
\centering
\includegraphics[width=11cm]{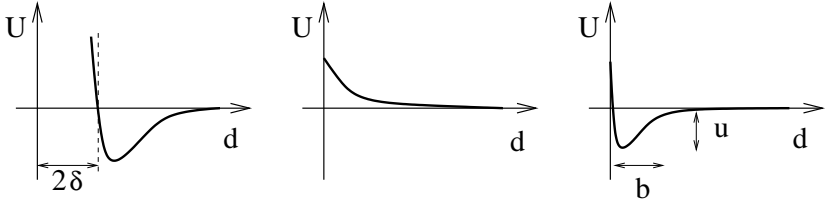}
\caption{Examples of pair interaction potentials achievable with model systems.}
\vspace{.0cm}
\label{fig2}
\end{figure}

From a fundamental point of view, playing with the set of interactions mentioned above,
one can to some extent tailor the interparticle potential, with control of both 
amplitude and range (see e.g Figure \ref{fig2}b,c,d),
and thus
prepare in the lab model colloidal systems.
In particular the Edinburgh group
has managed in the 80s to prepare colloidal suspensions that behave
as genuine hard spheres solutions \cite{pus1}. 
As such this substantiates the claim that
model colloidal systems constitute a rich tool-box for the study
of the collective behaviour that leads to crystallization or to the glass transition.
The reader is however warned that the mastering of the ``tunability'' 
of the interaction requires a significant expertise, especially in the 
chemistry and surface chemistry involved in the preparation
of the particles.

\subsubsection{More common colloidal systems}

The somewhat naive textbook picture for model colloids
given above must rapidly be supplemented by corrections that are often relevant:

{\em - Colloidal particles are not always rigid}. As such they can elastically
deform under internal or external forces, which renders the description of their interactions
more complex.
This in particular the case for emulsions and foams, where the dispersed phase
(the particles) is liquid or gaseous. When the system is concentrated
the particles pushed against one-another tend to facet at the cost of an elastically stored surface energy.
In other systems the elementary particles are elastic and compressible
in their bulk. This is for example the case for small gel beads \cite{bor1}
or multi-lamellar vesicles (onions) that can expel solvent and deform
upon compression \cite{rou1,ram2}.

{\em - Colloidal particles are not necessarily spherical.}
Rods or needle-shape objects are rather common,
as well as plate-like or disk-like particles. Such objects
start to interact and produce collective behaviour at volume fractions
much weaker than their spherical counterpart. For example laponite
systems are made of disk-like particles of a few nm thickness and about 100 nm
in diameter, that deviate form dilute behaviour at very low volume fractions \cite{bon1}.

{\em - Colloidal particles are seldom/never uniform.} This is in particular true for the surface charge:
heterogeneities in surface charge can lead to very anisotropic electrostatic interactions,
and even to attractions between particles of same {\em average} charge.
For example, laponite particles are said to have a different charge density on
their perimeter and on their faces for some pH and salt values, allowing for the building 
of ``house of cards'' structures.

{\em - Colloidal particles are seldom/never perfectly smooth}, in line with the two previous points. 
Their roughness can affect their interactions as short distances, control the apparition of capillary 
bridges in mixed solvents, and render the link between two particles in the primary minimum rigid 
\cite{rus1}.

{\em - Colloidal systems are seldom/never monodisperse.}
Large variations in size,  shape factor, etc.. is the rule for many practical systems.

{\em - There is rarely a single kind of particle in a colloidal system.} 
One extreme, industrial systems (creams, paints, etc ..), are often made of
at least 5 or 6 different main ingredients aimed at fulfilling different
(and often conflicting) tasks (see the notes of B. Cabane).

Eventually, while we are later in these notes going to discuss physical aging and a mild
form of history
dependence in these systems, a much more stringent one is that the physico-chemical state
of the system, even at small scales,
can in general not be anticipated from its sole composition.  
For example polymers adsorbing on the particles surface can lead to protected colloids
if added in a dilute suspension (a complete protecting layer forming around each particle),
but can induced bridging and thus promote aggregation if added when the particles 
are close to one another. Similarly any earlier stage at which the particles have been 
in contact can induce chemical reactions and irreversible sticking
between the particles that will resist later attempts to re-disperse the system.

At this stage, the second point of subsection 1.1. appears rather a matter of trust:
although clean model colloidal systems can be prepared,
many common colloidal systems rely on very varied and uncontrolled
interactions. 
As such it is not clear that 
such various and complex systems should obey rather universal features.
However we will see below that they share some similarity 
in terms of mechanical behaviour, which suggests that it is worth digging further,
if only to clarify in what sense ``glassy'' dirty colloidal systems 
differ from ``model glasses''.

\subsection{Gels or glasses: various kinds of soft solids ?}

Eventually let me stress that since the ``soft glass/concentrated colloid'' 
analogy has developed, some confusion
has invaded the community working on colloidal systems, that 
is often mirrored in the terminology used to describe the soft solids
that these systems form when concentrated.
The terms ``gel'', ``physical gel'', 
``glass'', ``paste'', ``jammed system'' and others are often used interchangeably,
with various authors obviously having different definitions 
of these terms in mind.
There is a great need for a clearer
classification (either from the point of view of underlying interactions
or resulting behaviour) and for a more systematic terminology, 
that hopefully the years to come could provide.

\begin{figure}[t]
\centering
\includegraphics[width=9cm]{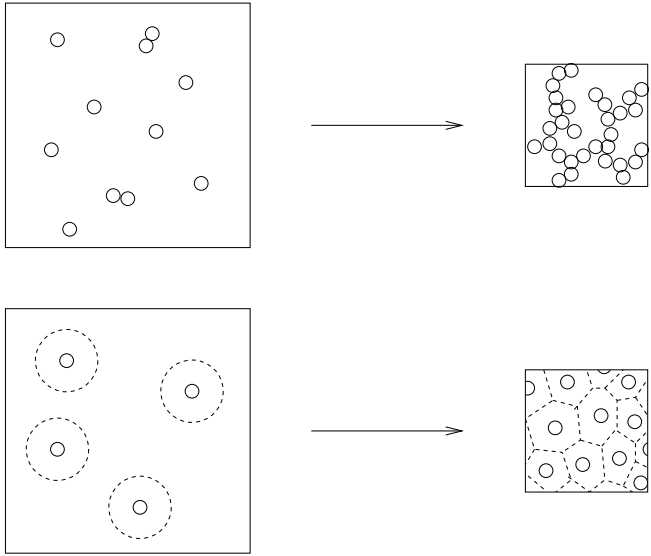}
\caption{Two scenarios for the formation of soft solids upon concentration. Top: sticky particles 
forming aggregates. Bottom: soft repelling particles under osmotic compression
(the dashed lines represents the range of the repulsive potential).}
\label{fig3}
\vspace{-.4cm}
\end{figure}

Schematically, there are two extreme pictures along which colloidal 
systems behave when concentrated, corresponding to
very different underlying interactions (see Figure \ref{fig3}).

The first one is that of rigid spheres that stick to one another upon contact, 
with a rather strong and rigid binding (e.g. due to local chemical binding, or because
the roughness of the surface
prevents any further rotation). 
Such particles diffusing in solution tend to form fractal aggregates
that also diffuse and aggregate, so that if the concentration is large enough,
one ends with a (history dependent) 
network made of rigidly interconnected aggregates. The volume fraction
of such a physical ``gel'' may be rather weak (sometimes only a few percent), 
with solvent filling the many voids in the aggregates structure (cite see e.g. \cite{lew1} and
references therein).

The second image is that of particles that repel through a soft potential
(either rigid particles with a repulsive potential of some range or
deformable particles with a short range repulsion) .
As concentration is increased (typically beyond
roughly 50 percent), the particles are ``pushed'' into one another,
the energy of the system increases and it is trapped in a metastable situation 
of disorderedly crowded particles\cite{clo1,der2}. This is what many authors call a soft glass.
In contrast with the previous rigidly connected 
``gel'', such a  ``glass'' has a positive osmotic pressure and 
should swell noticeably and unlimitedly upon addition of solvent.

Given the discussion in the previous subsection,
it should be clear that there is potentially a continuum of colloidal systems that connect
these two limiting cases (see for example the studies of the Weitz group
in the last decade where the attraction energy is systematically varied \cite{wei1},
and e.g. \cite{mou1,del1} and references therein). 
In most systems there are both attractions and repulsions,
the intensity of which is often comparable (and of order a few $k_bT$). A systematic 
``gel or glass'' distinction according to the interactions is difficult,
since it is not clear that the elementary object at the scale of which such a ``gel or glass'' 
distinction should/could be made is the individual colloidal particle rather
than some kind of aggregate.
Eventually,  the rheological properties of such systems are usually probed through dynamical schemes,
so that a static picture may not be sufficient to anticipate the response:
in some systems the weak aggregation of soft particles can lead to the formation of 
transient rigid aggregates that can resist a finite force (for a certain amount of time).

\subsection{Wrapping up the Introduction}

 To conclude, colloidal systems are very diverse. Model systems can be 
produced in the lab that may be used to study fundamental issues regarding the glass state and
the glass transition. Most practical systems behave when concentrated as
soft and weak solids 
that can be forced to flow under mechanical constraint.
These solids are in general {\em soft} as their elastic modulus is low (down to a few Pascal 
in some cases),
and {\em weak} because they flow plastically under weak stresses (also a few Pascal for some systems). 
The behaviour of these soft-solids
is however likely to be somewhat system-dependent, and a systematic classification of 
their behaviour in relation to their microscopic nature would be welcomed.\\

 The remainder of theses notes is devoted to experimental studies and modelisation
of these soft and weak solids, with an emphasis (especially in section 2)
on systems with little attractions (i.e. away from the strong gel picture). 
I will from now on very non-specifically
refer to those as ``soft-solids'', 
rather than using more specific names that could bear undesired meanings 
(paste, gel, glass, yield stress fluid).

\section{Experimental facts 1: soft solids that flow and age}
\subsection{Concentrating colloidal suspensions: from a ``viscous liquid'' to a ``soft-solid'' behaviour}

At low concentrations, colloidal suspensions tend to flow in a Newtonian way under
weak stresses. In particular at steady-state the shear stress $\sigma$ is linear in the shear rate
$\dgam$: $\sigma =\eta \dgam$ where $\eta$ is the viscosity of the suspension.
Often a shear-thinning regime takes over for stronger shear as evidenced by a 
bending in the stress-shear rate curve. The onset of non-linearities $\dot{\gamma}_{NL}$
is often identified with $1/\tau$
the inverse of the longest relaxation time of structures 
in the suspension.
In a related way, when probed through oscillatory rheology (see definitions
in the lectures of M. Cates), the system behaves as a viscous liquid $G''(\omega) \gg 
G'(\omega)$ at frequencies
well below $1/\tau$, and as an elastic solid at larger frequencies.

\begin{figure}[t]
\centering
\includegraphics[width=11cm]{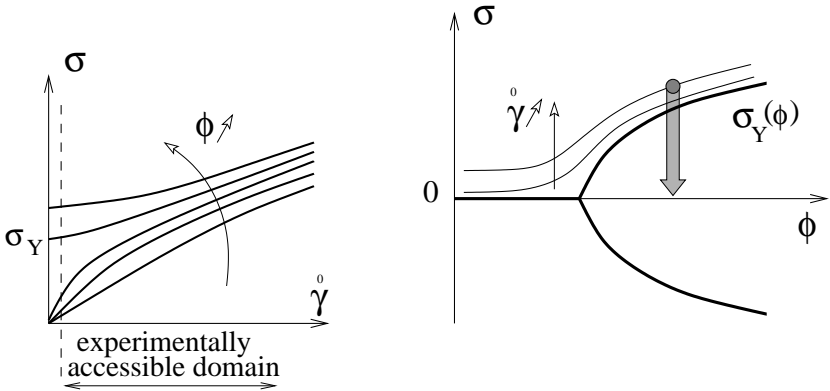}
\caption{Sketch of a typical steady-state flow response of a concentrated suspension.
Left: steady-state flow curves $\sigma(\dgam)$ for increasing volume fractions $\phi$.
Right: steady-state stress as a function of volume fraction $\phi$ for
different imposed shear rates $\dgam$ 
(the thick line corresponds to the 
limit curve for $\dgam \rightarrow 0$, i.e. $\sigma=\sigma_Y(\phi)$). 
The arrow is a schematic representation of
a quench from a mechanically fluidized system
into the soft-solid region.}
\label{fig4}
\end{figure}

Increasing the volume fraction, the viscosity of the system rises and the range of linear 
behaviour decreases. At even higher concentrations, the steady-state flow curve looks
altogether different. An often encountered situation
is  schematized in Figure \ref{fig4}: 
for low shear rates $\dot{\gamma}$, the stress tends to a finite value, the yield stress,
often denoted $\sigma_Y$ \cite{mas1,der2,clo1}.
At higher shear rates (in the non linear domain) the behaviour is however not
much different from what it looks like at weaker concentrations.
It is important to note that what happens at low $\dot{\gamma}$ is not easy to 
ascertain experimentally. Rheometers allow reliable measurements only 
down to a finite value of $\dgam$, so it is usually impossible to discriminate
between a power-law fluid with a weak exponent and an actual finite limit $\sigma_Y$.
A common engineer's formula to
encompass both is the Hershel-Buckley one: $\sigma \rightarrow \sigma_Y + A \dot{\gamma}^{\alpha}$
for $\dot{\gamma}\rightarrow 0$. In addition it may also be difficult to guarantee that 
a steady-state is indeed reached in such conditions.
\begin{footnote}
{
Note that the existence of a finite yield stress $\sigma_Y$ does not forbid
non-steady flow for a fixed weaker stress: creeping flows with 
a decreasing $\dot\gamma (t)$ are allowed. 
}
\end{footnote} 
  
Under oscillatory 
rheology the response becomes that of a {\em soft solid}
$G' \gg G''$ even at low frequencies.
Various frequency dependence of the moduli have been observed:
they can both increase with the same weak power-law, stay
both roughly constant, or
$G'$ can be roughly constant with $G''$ displaying a shallow minimum at rather low frequencies.\\

Obviously some kind of transition occurs upon increase of the concentration,
bringing the system from a viscous liquid state to a soft-elastic one.
The longest relaxation time-scale and the viscosity increase noticeably (diverge ?)
when the transition is approached. 
Conversely, effective values for the putative yield stress decrease on the solid side
when the concentration is decreased.
We will from now on deal only with the
soft-solid states.

\subsection{Probing the system in its ``soft-solid'' phase}

In the solid phase, the system appears as frozen
in some region of phase space 
given that a finite perturbation is required to
have it flow. In most cases however, its local structure is disordered 
and amorphous (only very clean systems crystallize). 
It is thus likely that such systems lie in a metastable state that is noticeably history dependent.
Consequently, to study the mechanical properties of these soft-solids,
it is first necessary to define reproducible preparation procedures. 
In analogy with
thermal quenches performed for molecular glasses or magnetic spin glasses, a natural way
is to prepare the system in a reproducible fluid initial state, so as to ``erase'' any
history dependence,
and then to quench it quickly in the soft-solid phase.

Unfortunately, changing the concentration or other physico-chemical parameters
homogeneously and rapidly throughout the sample is usually very difficult,
and temperature is not always very relevant here (unless in the vicinity of some 
structural thermodynamic transition, see e.g. \cite{ram2}).
An alternative is to take advantage of the flow induced fluidization of the system
to perform a mechanical quench \cite{der2,clo1,via1},
which is the route represented by the vertical arrow in Figure 4. 
A steady or oscillatory shear of strong amplitude is applied for some time to the system,
with the hope that this leads to a reproducible initial state. Cessation of flow then quenches the system 
in the soft-solid phase, with the history of the soft-solid starting at this ``initial time''
that we will call $t=0$. The validity of such a mechanical quench
should of course be asserted by experimental verification of the reproducibility.\\
  
\subsection{Mechanical aging}

An issue which has recently received attention,
fed by ideas and concepts coming from the community
working on the glass transition, is that of physical aging.
In the present context, we focus here on mechanical
aging, i.e. the influence of the age $t$ of the system since its quench
on its mechanical properties.
A measurement usually lasting some time, it is common to introduce the following notations
for the time sequence. The quench is at $t=0$. The system is left at rest
until $t=t_w$ (the waiting time), time at which a measurement is started.
If the system is not ergodic on experimental time scales,
the instantaneous outcome of the measurement at time $t$ depends on
the time $t'=t-t_w$ elapsed since the onset of the measurement (as is common),
but also in a systematic way on the waiting time $t_w$.

\begin{figure}[t]
\includegraphics[width=11cm]{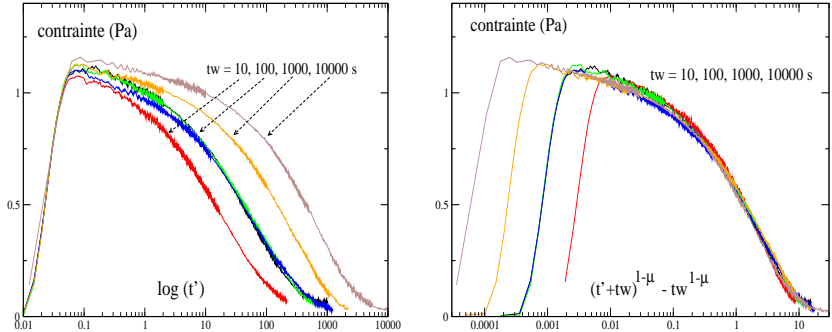}
\caption{Experimental stress relaxation measurements after different waiting times (linear-log plot)
for a solution of PEO-protected silica particles \cite{der4}:
(left) as a function of time $t'$ elapsed since onset of the deformation,
(right) as a function of the rescaled variable $(t'+t_w)^{1-\mu}-t_w^{1-\mu}$,
which is roughly $\simeq t'/t_w^{\mu}$ at times $t'$ shorter than $t_w$}
\label{figexp}
\vspace{-.1cm}
\end{figure}

Recent systematic experiments on microscopically
very different colloidal soft-solids have indeed
shown a strong dependence
of the mechanical relaxation time or time-scale $\tau$ of the system, on
its ``age'' $t_w$ (see e.g. fig. \ref{figexp}). Remarkably a clean power law relation is found in many cases
$\tau \sim t_w^{\mu}\tau_0^{1-\mu}$, with $\tau_0$ a system dependent time. The exponent $\mu$ 
is called the {\em aging exponent} in the glass community  \cite{str1,bou1}.
Let us cite a few examples to emphasize the great disparity of systems:
such a law has been found through stress relaxation measurement
(i) in a solution of PEO-protected spherical silica particles \cite{der2} with $\mu \sim 0.5-0.7$,
(ii) in a dense suspensions of onions made 
of surfactant bilayers ``doped'' with copolymers \cite{ram2} with $\mu \sim 0.7-0.8$,
and through creep measurements (iii) in dense suspensions of swollen micro-beads
of polyelectrolyte gel \cite{clo1} with $\mu$ close to 1.
For other systems the dependence on the age of the system can be even stronger:
the effective oscillatory viscosity of laponite solutions
was found in \cite{bon3} to increase exponentially with $t_w$
which suggests $\tau \sim \exp(t_w/\tau_0)$ (i.e. $\mu \rightarrow \infty$).\\

This demonstrates physical aging through a systematic route in these systems: 
the dynamics of the system gets slower and slower as time goes on,
with less stable arrangements more rapidly disappearing leaving the system
in deeper and deeper metastable states. This slow drift 
can in principle be driven by energy stored in the system,
by experimental mechanical agitation or noise, or by genuine thermal agitation.
The description of the mechanisms at work in this 
slow evolution of the system are clearly the clue to a proper 
and complete understanding of the physics of aging.
In the following we by-pass this question for the moment
and report  on a recent attempt to build
a simple phenomenological model able to account for the
observations quoted in this section. A more explicit and sophisticated
but still phenomenological set of equations,
the SGR model, is presented in the lectures of M.E. Cates.

\section{A class of simple models}
\subsection{A Maxwell model with one scalar internal variable}

Essentially the aim is here to write down the simplest model that accounts
for the observations reported above for the ``soft-solid state'':
(i) there is no obvious intrinsic rheological relaxation time, so that the instantaneous relaxation
spectrum ages progressively, (ii) shear is able to fluidize the system 
and to bring it to a flowing steady-state with a finite instantaneous relaxation
spectrum. We implement these features in a model that neglects for the moment
heterogeneities.

This can be achieved through a simple adaptation
of the Maxwell model, which is the simplest visco-elastic model, with a single time scale 
$\tau$, corresponding to an elastic element in series with purely viscous one.
The resulting equation for a scalar description
of shear linking the stress $\sigma$ to the applied shear rate
$\dot{\gamma}$ is then: $\partial_t\sigma=-\sigma/\tau + G_0\dot{\gamma}$,
with $G_0$ the elastic modulus and $\eta=G_0\tau$ the viscosity.
For the soft-solids that we aim to describe there is no intrinsic time scale,
but an instantaneous one that drifts towards infinity spontaneously, 
unless an applied shear fluidizes the system.
We thus chose to take the mechanical relaxation time scale as a variable describing the instantaneous
state of the system, the evolution of which is ruled by the competition of 
spontaneous aging and flow induced fluidization.
We use equivalently $a=1/\tau$, which we call the
``fluidity''of the system. The resulting set of equations read:
\begin{equation}
\partial_t\sigma=-a(t) \sigma  + G_0\dot{\gamma}
\label{e1}
\end{equation}
\begin{equation}
\partial_t a= - f(a) + g(a,\sigma,\dot{\gamma})
\label{e2}
\end{equation}
where $f$ (spontaneous aging) and $g$ (shear-induced fluidization)
are positive functions. Aging requires that under the sole action of $f$,
$a$ tends towards $0$ as time goes on. $G_0$ is taken constant in the simplest picture.

For long waiting times and weak shear the fluidity is expected to be small (the relaxation time is very long),
so what will matter is the behaviour of functions $f$ and $g$ in the vicinity of $a \simeq 0$.
Using rather formal expansions we write:
\begin{equation}
f(a)_{a \rightarrow 0^+}\,  \simeq r_1 a^{\alpha} + r_2 a^{\alpha +\beta} + ...
\label{e3}
\end{equation}
\begin{equation}
g(a,\sigma,\dot{\gamma})_{a \rightarrow 0^+} \simeq \,  u_1 a^{n_1}\sigma^{m_1}\dot{\gamma}^{p_1}
 + u_2 a^{n_2}\sigma^{m_2}\dot{\gamma}^{p_2} + ...
\label{e4}
\end{equation}
It is expected that $g$ should be even in ($\sigma$,$\dot\gamma$),
so that if $g$ is analytic then $m_1+p_1$ and $m_2 +p_2$ should be even. If $g$ is not analytic,
then $\sigma$ and $\dot\gamma$ in the above equation should be replaced by their absolute values.

In our earliest attempts we proposed in an analogy with a Landau description
of a transition where $a$ would be the order parameter, to consider that 
$r_1$ increases with concentration from a negative value in the viscous-liquid 
state to a positive one in the soft-solid one: $r_1 \sim (\phi-\phi_c)$.
We will not pursue this any further here (see \cite{der3,der4} for related discussions) 
and instead focus on the properties of the soft-solid phase,
taking $r_1$ and $u_1$ as positive constants.

\subsection{Relation to other models in the literature}
\subsubsection{Bond models for gels}

Given that it is the simplest approach, the use of a single scalar parameter
to describe the state of the system has been pursued in many works.

For example an empirical model for particles that stick to one another
describes the instantaneous state of the system by an average number of bonds per particle 
$\Lambda$. This number tends to increase spontaneously, but is reduced by the action of
shear breaking some bonds.
A set of equations proposed by Coussot et al. in \cite{cou2}
is $\partial_t \Lambda= 1/\tau - A \Lambda \dot{\gamma}$, complemented 
with the phenomenological rheological law $\sigma/\dot{\gamma}=\eta=\eta_0(1+\Lambda^n)$,
with $\tau$ and $A$ constants, and $n$ an exponent.

Taking the rheological law to be roughly equivalent to the Maxwell equation (\ref{e1})
if stress relaxes faster than $\Lambda$, this model can be more or less recast into (\ref{e1})(\ref{e2})
with $a= (G_0/\eta_0)(1+\Lambda^n)^{-1}$.
In that case, the exponents proposed in (\ref{e3})(\ref{e4}) read:
$\alpha=1+1/n$, $n_1=1$, $m_1=0$, $p_1=1$.

\subsubsection{Activated hopping}

\begin{figure}[t]
\centering
\includegraphics[width=9cm]{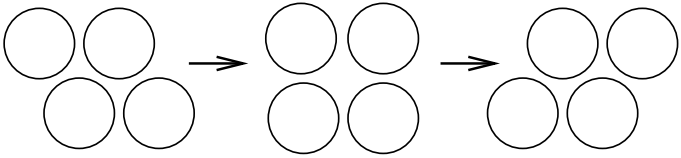}
\caption{Elastic deformation and rearrangement
under shear of a set of particles}
\label{fig6b}
\end{figure}

In many models for plastic flows of amorphous solids,
flow is thought to occur by the elastic distortion of ``elements''
up to a point where they become unstable to a rearrangement
of the local structure (see e.g. \cite{liu2,deb1,sol1}),
rearrangement that yields a reorganization of the stress field 
on other elements (Figure \ref{fig6b}).

For these plastic events to occur, an activated state must thus be reached,
which is usually a conformation of higher energy
and lower density (to allow the particles to pass each other).
It is often chosen to describe the rates for such events
as the exponential of an activation factor $\cal E$ that
is often conceptualized as $\Delta E /T_{\rm eff}$ with $\Delta E$ the barrier to overcome and 
$T_{\rm eff}$
the effective temperature in the system \cite{sol1}.
The rheological equation is then $\dot{\gamma}=\dot{\gamma}_{\rm elastic} +
\dot{\gamma}_{\rm plastic}$, with $\dot{\gamma}_{\rm elastic}=\partial_t\sigma/G_0$
and in a linearized version $\dot{\gamma}_{\rm plastic}\simeq \tau_0^{-1}e^{-\cal E}(\sigma/\sigma_0)$
as in the Eyring model ($\tau_0$ and $\sigma_0$ are constants).

This is very similar to our adaptation of the Maxwell model with $a=(\sigma_0\tau_0)^{-1} e^{-\cal E}$.
If the effective temperature of the system decreases as the system ages, or conversely
if the typical barrier for rearrangement increases then this should be described
by an evolution equation for $\cal E$. 
This spontaneous evolution itself is likely to occur through activated rearrangements
that may or may not be identical to those
probed by the shearing. This yields for the spontaneous
evolution of $\cal E$ a law of the kind $\partial_t {\cal E} \sim K e^{-n{\cal E}}$
with $K$ a constant and $n$ the ratio between the activation factor
for these rearrangements and those corresponding to the plastic shear deformation
(if those are identical, as e.g. in the SGR model \cite{sol1} then $n=1$).
The above mentioned dynamics for ${\cal E}$ corresponds to an exponent $\alpha= 1 +n$
in (\ref{e3}).

In a similar description inspired by a ``free-volume'' picture for activated rearrangements,
where ${\cal E} =\Delta v/v_{\rm free}$ and the free-volume $v_{\rm free}$ measures the compaction 
of the system and controls the statistical likelihood of an activation volume $\Delta v$,
Lemaitre \cite{lem1} has recently proposed a model that in its linearized isotropic version
boils down to something very close to equations (\ref{e1}-\ref{e4})
(with additional logarithms) with $\alpha=1+n$, $n_1=2$, $m_1=2$, $p_1=0$.

\subsubsection{a 4-state model}

Eventually let me briefly mention a model that we proposed a few years 
ago inspired by the same physics than in the previous subsection 3.2.1 \cite{der1}.
In this model each element is supposed to be in one of the four following state of stress
$-3, -1, +1$ and $+3$. The dynamics of the system (the evolution of the probability for
an element to be in a given state of stress)  is ruled by three processes: (i)
a positive shear rate $\dot{\gamma}$ tends to shift the elements to states
of higher stress,
(ii) the states $-3$ and $+3$ are unstable, so that they rearrange and 
relax with a decay rate $1/\tau_0$
to either of $-1$ or $+1$, (iii) these rearrangements trigger a random re-shuffling of the
stresses of other elements modeled by an effective diffusion of the stress value 
at a rate $D$ proportional to the instantaneous number of rearrangements.
The dynamics of this system was shown to be similar to (\ref{e1}-\ref{e4}) in the vicinity of
the liquid/soft-solid transition observed, with $D$ proportional to the fluidity $a$,
and $\alpha=1$, $\beta=1$, $n_1=0$, $m_1=0$, $p_1=1$.

\subsubsection{a concluding remark}

In conclusion, the simple class of models described by equations
(\ref{e1}-\ref{e4}) roughly encompass various models in the literature
corresponding to specific choices for the phenomenological exponents
describing the behaviour of $f$ and $g$ at low
fluidity. We will consequently pursue without much restrictions on the latter to 
examine the behaviours described by these models.
In particular, $\alpha$ need not be an integer.
Let us however point out that in all the models quoted here $\alpha \ge 1$,
which simply states that at long waiting times (low values of $a$),
the instantaneous rate of decrease of $a$, which reads $r_1 a^{\alpha-1}$, decreases
with time (in line with the picture of physical aging).
We thus restrict ourselves to $\alpha \ge 1$ in the following.

We will now examine in some detail the predictions of this model, leaving a discussion
of its virtues and shortcomings to section 5.

\subsection{Predictions}

Let us now briefly span some of the predictions of the model for various mechanical probing.
We refer the reader to \cite{der3,der4} for a more thorough presentation. 
In all cases the system has been quenched at $t=0$, and the measurement 
starts at $t_w$.

\subsubsection{Mechanical aging}

An immediate remark  is that during the spontaneous relaxation before the 
onset of the measurement,
$a$ decreases with time. 
More precisely $a(t) \sim 1/(r_1t_w)^{\mu}$ for long times
with an aging exponent:
\begin{equation}
\mu=1/(\alpha-1)
\end{equation}
This holds for $\alpha > 1$, with for the marginal case $\alpha=1$,
 $a(t) \sim \exp(-r_1t_w)$.

We show below that $\mu$ indeed coincides with the definition
given in 2.3: the characteristic mechanical time scale of the system $\tau$
is the inverse of the fluidity at the onset of this measurement, and 
$\tau \simeq \tau_0^{1-\mu} t_w^{\mu}$ 
(with the microscopic time $\tau_0= r_1^{\frac{\mu}{1-\mu}}$).

An important distinction can be obtained by inspection of
equation \ref{e1} to \ref{e4}: if $\alpha >2$ 
then $a$ relaxes much slower than the stress $\sigma$,
whereas 
if $\alpha < 2$, $a$ relaxes to zero faster which freezes the dynamics 
and prevents $\sigma$ to relax down to $0$. 
This distinction matches here the classical one between 
subaging ($\mu <1$) and superaging ($\mu >1$).
``Full aging'' ($\mu=1$) here corresponds to the limit case $\alpha=2$,
where the same processes rule the relaxation of $a$ and $\sigma$.

Let us now assert these points by examining
the linear  response to a stress-relaxation measurement performed
at $t_w$: at $t_w$ a weak deformation of amplitude $\gamma_0$ is applied to the system
and then maintained fixed. This creates an almost instantaneous stress $\sigma=G_0\gamma_0$
that subsequently relaxes as $\sigma(t)= G(t',t_w)\gamma_0$,
where $G(t',t_w)$ is the time-dependent modulus that for an aging system depends on both
the age of the system at onset of the deformation $t_w$ and of the time elapsed since onset $t'=t-t_w$.

\begin{itemize}
\item for $\alpha >2$ (i.e. sub-aging $\mu <1$), 
\begin{equation}
G(t',t_w)=G_0\exp
\left[ -
\frac{(t'+t_w)^{1-\mu}-t_w^{1-\mu}}{(1-\mu)\tau_0^{1-\mu}}
\right]
\label{eq6}
\end{equation}
This kind of scaling variable appears in the description of mean field 
models for glassy phases in spin-glass systems \cite{bou1}.
For times shorter than the system's age $t'\ll t_w$,
$G(t',t_w) \sim 
\exp[-t'/\tau]
$,
a simple exponential relaxation with a characteristic time $\tau =\tau_0^{1-\mu}t_w^{\mu}$. 
For much longer times $t'\gg t_w $,
$G(t',t_w) \sim \exp[- t'^{1-\mu}/( (1-\mu)\tau_0^{1-\mu})]$,
the stress decreases to $0$ as 
a stretched exponential 
\begin{footnote}
{
Note that the stretched exponential here stems from the non-linear equation,
in contrast to the classical picture for glassy systems of a convolution
of single exponential processes with a distribution of time scales.
} 
\end{footnote}
with a stretching exponent $1-\mu$ smaller than $1$.

\item for $\alpha <2$ (i.e. super-aging $\mu >1$), 
\begin{equation}
G(t',t_w)= G_0 \exp
\left[ -
\frac{
\frac{1}{t_w^{\mu-1}}- \frac{1}{(t'+t_w)^{\mu-1}} 
}
{(\mu-1)\tau_0^{1-\mu}}
\right]
\end{equation}
At short times $t' \ll t_w$
we have again
$G(t',t_w) \sim \exp[-t'/(\tau_0^{1-\mu} t_w^{\mu})]
$,
whereas at longer times 
$t'\gg t_w $, relaxation proceeds as
$G(t',t_w) \rightarrow G_{\infty}(t_w)= G_0\exp(-(\tau_0/t_w)^{1-\mu})$,
i.e. there remains a non-zero residual stress $\sigma_{\infty}=G_{\infty}(t_w)\gamma_0$
that increases with $t_w$.

\item for $\alpha=2$ (full aging $\mu=1$), The stress decays algebraically:
\begin{equation}
G(t',t_w)\simeq G_0 \left(\frac{t_w}{t'+t_w}\right)^{1/r_1}
\end{equation}

\end{itemize}

\subsubsection{Non-linear ``steady-state'' rheology}

We now address the most common rheological characterization,
namely the steady-state flow curve $\sigma(\dot\gamma)$.
With the equations at hand, such a steady-state is always reached if a
fixed $\dot\gamma$ is applied, with the history of the system erased
(aging suppressed).
Using (\ref{e1}), the fluidity $a$ in this steady state is set by $f(a)=g(a,G_0\dgam/a,\dgam)$.
For low values of $\dgam$, the fluidity remains weak so that using the formal
expansions (\ref{e3},\ref{e4}), one finds from the competition
of spontaneous aging and flow-induced rejuvenation
$a \sim \dgam^\frac{\nu-\epsilon}{\nu}$,
which leads to a power law steady-state relation 
$\sigma (\dgam \rightarrow 0)\sim \dgam^\frac{\epsilon}{\nu}$, 
where $\nu =\alpha-n_1+m_1$ and $\epsilon=\alpha-n_1-p_1$.
Note that $\nu-\epsilon=m_1+p_1$
is likely to be a positive number if shear is to enhance fluidity in (2)-(4).

\begin{figure}[h]
\centering
\includegraphics[width=9cm]{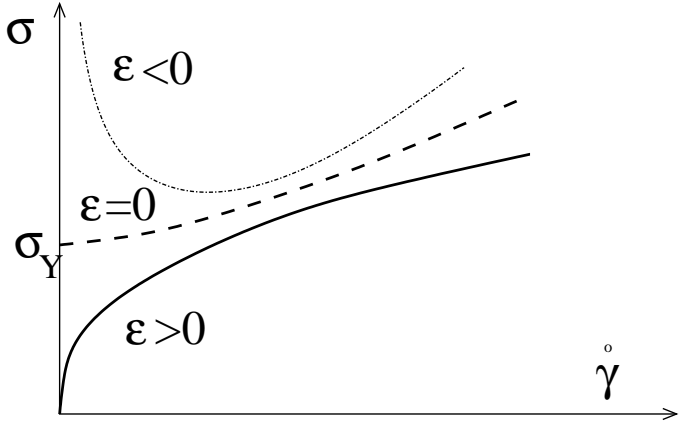}
\caption{Steady-state rheology
as a function of the exponent $\epsilon=\alpha-n_1-p_1$. The case $\epsilon<0$ is unphysical (see text).
}
\vspace{-.3cm}
\label{fig6bis}
\end{figure}

So the low shear steady-state behaviour leads us to the following distinction
according to $\epsilon=\alpha - n_1 - p_1$ (see Figure \ref{fig6bis}):

\begin{itemize}
\item for $\epsilon >0$ , at low shear rates the fluid behaves as a power law fluid
$\sigma (\dgam \rightarrow 0)\sim \dgam^\frac{\epsilon}{\nu}$ (full line in Fig.  \ref{fig6bis}),

\item for $\epsilon =0$, stress tends to a finite value $\sigma_y=G_0^\frac{p_1}{\nu}
(r_1/u_1)^\frac{1}{\nu}$ (dashed line in Fig.  \ref{fig6bis}),

\item $\epsilon <0$ corresponds to a stress diverging as $\dgam$ goes to zero
which is unphysical for the local flow rule of a realistic system, which has clearly
finite stress limits for yield to occur.
\end{itemize}

For larger values of the shear, the next terms in the expansion of both 
$f$ and $g$ start to play a role, modifying the effective stress-strain rate behaviour
(see e.g. \cite{der3}). This can lead to some subtleties for the case $\epsilon =0$ as we will see later.\\

\subsubsection{More complex topics: oscillating rheology and transients}

This subsection deals with more intricate points that the reader may want to skip
in a first reading, if only to secure that it does not become the last reading too ...

{\em ``Steady state'' oscillatory rheology -}
An oscillatory measurement is performed on the pasty phase,
e.g. a deformation $\gamma=\gamma_0 \cos(\omega t +\psi)$ is applied from
$t_w$ on.
No matter how small $\gamma_0$, at long times it is this shear that
will control the oscillating value of the fluidity
through equation (\ref{e2})
(this is for example in contrast with the SGR model where
the effective temperature is fixed to a finite value, which guarantees an intrinsic linear domain).
As a consequence the long-time response is intrinsically non-linear
with many harmonics in the stress response.

\begin{figure}[t]
\centering
\includegraphics[width=9cm]{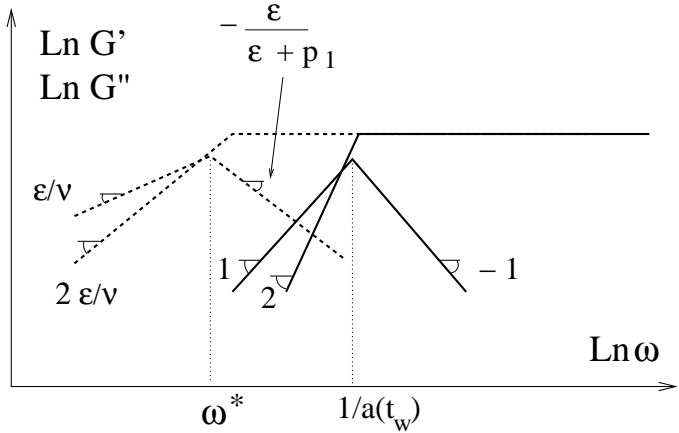}
\caption{Oscillatory rheology: $\gamma =\gamma_0 \cos(\omega t)$. 
Left (dashed lines): effective moduli in the asymptotic long-time regime 
(non-linear). Right (full lines): linear response at short times ($t-t_w \ll t_w$). As time goes on,
the measured moduli drift from the latter to the former. 
$\omega ^* \sim \gamma_0^{\nu/(\nu-\epsilon)}$ depends on the 
measurement amplitude}
\label{fig7}
\end{figure}

Although there are consequently no linear moduli $G'(\omega)$ and $G''(\omega)$,
we can still discuss the outcome of a rheometer measuring the in-phase and out-of-phase
first harmonic of the stress divided by $\gamma_0$, and call these
numbers $G'_{\infty}(\omega,\gamma_0)$ and $G''_{\infty}(\omega,\gamma_0)$ by extension
(the $\infty$ subscript is to emphasize that these are asymptotic long-time values).
A qualitative analysis developed in \cite{der3,der4} shows that one expect
a frequency dependence reminiscent of the Maxwell model in gross features
but with different slopes for $\epsilon >0$ (see dashed lines in Figure \ref{fig7}). 
The cross-over frequency $\omega^*$ in this
non-linear situation is amplitude dependent and weak if $\gamma_0$ is small.
Therefore experimental observations may well be restricted to the $\omega >\omega^*$ domain.
\begin{footnote}
{
The case $\epsilon =0$ yields non-linear constants
 $G'_{\infty}(\omega,\gamma_0)= G_0$ and
$G''_{\infty}(\omega,\gamma_0)\sim G_0 \gamma_0^\frac{\nu}{p_1}$.
}
\end{footnote}

{\em Transients and thresholds for non-linear behaviour}

The results mentioned above, both for steady and oscillatory rheology,
correspond to the long-time behaviour in measurement processes
by which energy is constantly fed into the system 
(in contrast 
to stress relaxation measurement - see 3.3.1 -, where a finite perturbation 
is made on a system of initially non-zero fluidity).

However, when the measurement (applied constant shear rate or oscillatory shear) 
starts at $t_w$
the system initially responds as a Maxwell fluid of characteristic time $1/a(t_w)\sim t_w^\mu$.
There is thus in the two experiments a cross-over in time
from the linear response at short times to the non-linear ``steady'' response at long times.
The duration of the linear response decreases if the amplitude 
$\dot\gamma$ or $\gamma_0$ increases, but also if the waiting time $t_w$ increases \cite{der3}.

Following the response in time, one thus anticipates in the steady shear experiment
an instantaneous viscosity that decreases from 
${\eta (t_w)} \sim 1/a(t_w)$
to its asymptotic value 
$\eta_{\infty} \sim \dgam^\frac{\epsilon-\nu}{\nu}$ for very weak applied shear
rate $\dgam$.
\begin{footnote}
{
If the measurement is so strong that it ``immediately'' leads the system 
into non-linearity, other features can be observed such as a peak (overshoot)
in the stress response to a fixed applied $\dgam$.
The value of this peak drifts with
the age of the sample, providing a simple way to assess physical aging in the system
(see \cite{der4} for a related discussion).
}
\end{footnote}

Similarly, if one follows the first harmonic of the stress in an oscillatory experiment,
one expects that the hump in $G''$ slides progressively to lower frequencies before saturating
at $\omega^*$ with the slopes of the moduli becoming progressively weaker (Figure \ref{fig7}).\\

\subsection{Intermediary conclusion}
The class of models described by the simple equations (\ref{e1}-\ref{e4})
generates a rich physics not dissimilar
to experimental observations quoted in section 2.
For systems under weak shear, the response 
is dictated by the behaviour for small fluidities of the functions
$f$ and $g$ describing spontaneous aging and flow induced rejuvenation 
in ($\ref{e2}$). If these behaviour are power-law like ($f \simeq r_1 a^\alpha$,
$g \simeq u_1 a^{\alpha-\epsilon} \sigma^{\nu-\epsilon}(\dgam/a\sigma)^{p_1}$),
then (i) physical aging is characterized by the exponent $\mu = (\alpha-1)^{-1}$
(see 3.3.1) and (ii) the low shear steady-state rheology is that of a power-law fluid for $\epsilon >0$
and of a yield stress fluid for the smallest acceptable value $\epsilon=0$.
We will see later that this last case deserves more careful attention.

\section{Experimental facts 2: soft solids that flow in a strange way}
\subsection{Avalanches and ``viscosity bifurcation''}

The rheological behaviour of colloidal soft-solids is sometimes  
less smooth than suggested by the flow curves discussed in the previous section.
A spectacular manifestation is that of surface avalanches
as reported in a recent letter\cite{cou2}:
a plane covered by a layer of soft-solid is
progressively inclined with respect to horizontal.
For a power-law fluid (e.g. flow curve akin to the lower curve in Figure \ref{fig6bis})
one expects a progressively increasing downhill sliding as the
slope is increased. For a yield-stress fluid 
(e.g. steady flow curve as the middle one in Figure \ref{fig6bis}),
one expects no flow before the tangential component 
of the stress reaches the yield stress value $\sigma_Y$, and a progressive flow thereafter.
What is in contrast sometimes observed, for a wide variety
of soft solids of very different chemical nature,
is the sudden onset of an ``avalanche''
for a stress $\sigma_c$, with some material spontaneously accelerating \cite{cou2}.

In more conventional rheometry, a related experiment is as follows \cite{cou2}:
for a given reproducible preparation scheme, apply at time $t_w$ 
a fixed stress $\sigma$ and subsequently measure the instantaneous
flow rate $\dgam(t)$ or equivalently an
instantaneous viscosity $\eta(t)=\dgam(t)/\sigma$. The observation, again for many complex systems 
of different microscopic nature, is that for
$\sigma$ smaller than a value $\sigma_c$, the viscosity $\eta(t)$ tends to infinity
with time (in the model terms the fluidity tends to $0$, the system freezes),
whereas for $\sigma$ larger than $\sigma_c$, the viscosity tends at long times to a finite value.
The long time viscosity $\eta(t \rightarrow \infty)$ actually jumps from $\infty$ to
a finite value at $\sigma_c$, hence 
the term ``viscosity bifurcation'' coined for this jump.

\begin{figure}[t]
\centering
\includegraphics[width=9cm]{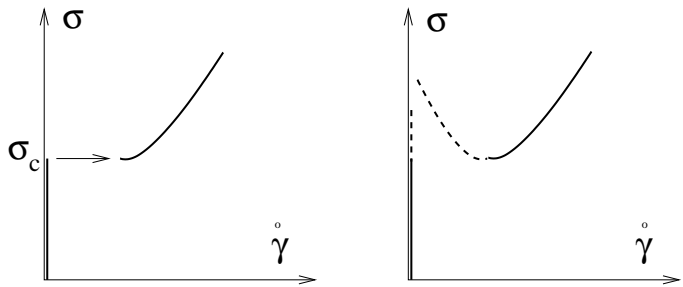}
\caption{Scheme for the observed viscosity bifurcation 
}
\vspace{-.3cm}
\label{fig8}
\end{figure}

Such a phenomenon suggests a jump between two branches
in the steady-state flow diagram (Figure \ref{fig8} left),
which in a simple picture can be the outcome
of an underlying flow curve with a decreasing branch (Figure \ref{fig8} right) .
 
\subsection{Parallel with flow induced transitions: heterogeneous ``banded'' flows}

In many complex fluids, non monotonic local flow curves are taken to be often responsible 
for transitions between
a high viscosity branch and a low viscosity one. 
Such a decreasing branch is notoriously unstable when inertial effects are taken into account.
As a result the system can separate into domains (bands)
of material corresponding to each of the stable branches.
Many studies have recently explored these flow induced transitions and started 
to shine some light on these complex far from thermal equilibrium problems
\cite{olm1,olm2,rad1,dho1,britt,ram1,berret1,diat}.
The resulting macroscopic flow behaviour, relating the macroscopic stress $\Sigma$ 
to the macroscopic or effective shear rate $\dGam$,
can be rather complex, with selection rules in the banded regime very
sensitive to gradient terms in the equation that decide of the stability of the interfaces
between bands \cite{lu1}.

Direct observation of banding 
in the slow flows of some soft-solids have been reported
\cite{pig1,cou3,clo2,deb1} and also shows up in some numerical simulations
\cite{deb1,bar1}. In both experiments and numerical work, boundaries where sometimes
shown to affect the banding phenomena.
One in principle expects all the complexity revealed by former studies 
on ``flowing'' complex fluids to be potentially present here, with in addition the specificities
linked to having one of the ``phases'' being a non-ergodic aging soft-solid.

\subsection{Description within the simple class of models}

\begin{figure}[t]
\centering
\includegraphics[width=11cm]{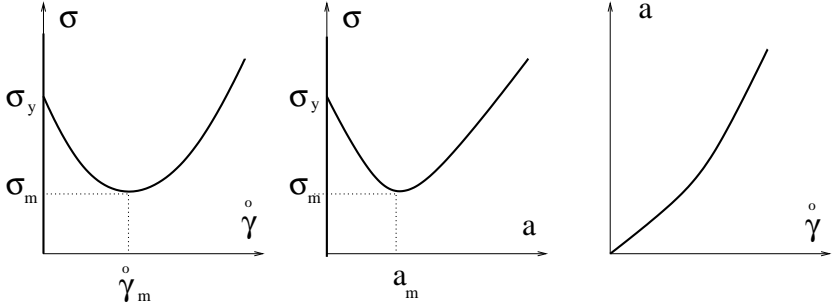}
\caption{Steady-state solutions of equations (3.1-3.2) for $\epsilon=0$ and a non-monotonic
$\sigma(\dgam)$.
}
\label{fig10}
\end{figure}

Given this potential complexity it is tempting to investigate such behaviour within 
our simple model by considering situations where the competition between 
aging and flow induced fluidization results in a local flow curve
similar to Figure \ref{fig8}b . 
A necessary condition within our model is that $\epsilon=0$.
 For example if $n_2+p_2 > \alpha +\beta$ and if $r_2$ and $u_2$ are positive,
then $\sigma _{\dgam \rightarrow 0^+} \simeq \sigma_y[1- (..) \dgam^{n_2+p_2 - \alpha}]$.
In such a case, both the $\sigma(a)$ and $\sigma({\dgam})$ plots present a decreasing 
section down to a value $\sigma_m$, before increasing monotonically again (Figure \ref{fig10}).

\subsubsection{Adapting the model}

We have investigated the behaviour 
of the system in this situation for the simplest geometry where heterogeneity
is allowed only in the direction of the velocity gradient (gradient banding): the fields $\sigma$,
$a$ and $\dgam$ vary along a single direction $z$ normal to the two boundaries in relative motion.
We then add a diffusion like term for the equation describing the evolution of
the local fluidity. I will not enter here the detailed description
of the model (the reader is referred to \cite{pic1})  
but rather focus on a few remarks before presenting the resulting macroscopic 
behaviour under steady driving, either at fixed stress $\Sigma$ or at fixed
strain rate $\dGam$.

- There is now an absolute need for boundary conditions on the walls, 
not only for the fluid velocity but also for the internal variables,
here the fluidity $a$. How the wall affects the density, degree of order and mobility of
a colloidal fluid in their neighborhood is a difficult
question, that will be shown to have important consequences on the overall
macroscopic behaviour.

- The choice of a diffusion term in the equation for the evolution of
$a$ is debatable, and used here as a 
simple choice involving gradients. Obviously the
corresponding diffusion coefficient $D$ may itself depend on $a$. 
As a result the thickness $\ell$ of the interface between frozen ($a \sim 0$) and fluidized
regions ($a$ finite) in a banded flow, is fixed on the frozen side by the competition
between $f(a)\sim - r_1 a^{\alpha}$ and the diffusion term scaling as $D(a) a/\ell^2$,
which yields $\ell \sim (D(a)  a^{1-\alpha})^{1/2}$ indicating an algebraic decay
unless $D$ scales as $a^{\alpha-1}$.
Thus in contrast with flow induced transitions between two branches of finite viscosity,
the interfaces between bands need not have a finite thickness, and the bands can
interact through algebraic effective potentials.

\subsubsection{Fixed stress $\Sigma$}

In the case where the macroscopic stress $\Sigma$ is maintained at a fixed value, 
within our model
the dynamics of the field $a(z)$ 
is such that the system evolves toward the local minimum of
a $\Sigma$ dependent functional (very similar to a free-energy in a field theory for wetting).

\begin{figure}[t]
\centering
\includegraphics[width=11cm]{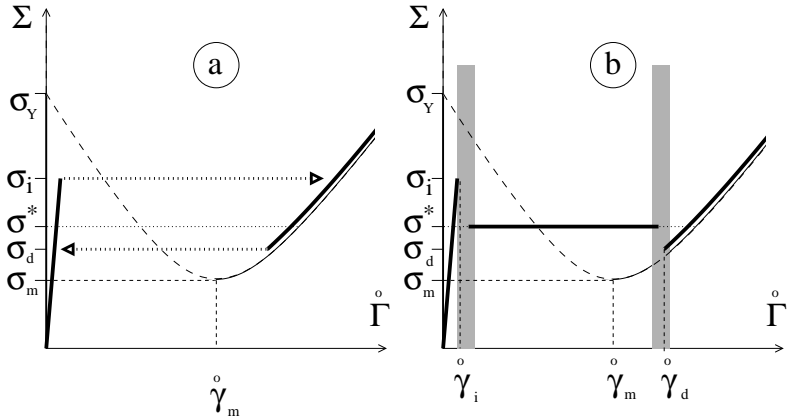}
\caption{Flow behaviour (full lines) under steady driving: (a) at fixed stress $\Sigma$, (b) at fixed 
average shear rate $\dot\Gamma$. The thin lines are the local steady-state solutions of (3.1-3.2)
depicted in Figure 10. Oscillating stress response can be observed in the grey areas.}
\label{fig11}
\end{figure}

Essentially there are two effective branches:
a frozen branch corresponding to a frozen bulk $a\sim 0$ with layers of finite
fluidity close to the wall induced by the boundary conditions,
and a fluidized branch with the bulk at a value of $a$ corresponding to the stress $\Sigma$
on the up-rising part of the local flow curve (Figure \ref{fig11}a) 
with layers of possibly weaker fluidity close to the walls.
It is important to point out that $\sigma_i$ and $\sigma_d$ in Figure \ref{fig11}a 
depend sensitively on
the boundary conditions for $a$ on the wall, as the walls can act as 
nucleation centers for either of the frozen and fluidized ``phases''.
\begin{footnote}
{
 A finite amount of noise in the dynamics 
can also trigger this nucleation in the bulk which would result
in the narrowing of the hysteresis loop around a value $\sigma^*$. 
}
\end{footnote}

For stress values $\Sigma$ lower than $\sigma_d$ only the frozen branch is stable,
whereas for values larger than $\sigma_i$ only the fluidized branch is.
For intermediate values
of the stress $\sigma_d<\Sigma<\sigma_i$,  both branches are possible solutions at long times.
{\em Which branch is selected depends then on the initial conditions}.
For a given preparation scheme of the system, and test under increasingly large applied fixed stresses,
one thus anticipate a sudden jump from the frozen to the fluid branch
at a given value $\Sigma_c$ that depends on the preparation scheme
(with $\sigma_d<\Sigma_c<\sigma_i$).
This is in agreement with the experimental picture for the ``viscosity bifurcation'' (section 4.1),
and further suggests that this value increases with the age of the system,
the transition occurring later for a weaker initial fluidity.

If the stress is ramped up and down slowly a hysteresis loop is described 
as indicated by the arrows on Figure \ref{fig11}a. In an avalanche experiment as
reported in 4.1 \cite{cou2}, $\sigma_i$ would thus play the role of the critical stress $\sigma_c$.
Other rheological experiments could conclude that the system has an effective
 yield stress anywhere between $\sigma_d$ and $\sigma_i$.

\subsubsection{Fixed shear rate $\dGam$}

The case of an imposed macroscopic constant shear rate $\dGam$ (i.e. the two bounding
plates moved at a constant relative velocity),
is much more complex as both the field $a(z,t)$ and $\Sigma(t)$
evolve in a coupled way.

The resulting macroscopic picture is represented on Figure \ref{fig11}b.
Basically the two branches described in the previous subsection are still present,
but for intermediate shear rates, the system can now stabilize in a banded state 
at a stress very close to $\sigma^*$. In these banded situations
the system is essentially divided into a fraction 
of the system at a fluidity close to $0$ and a fraction at a fluidity close to $a(\sigma^*)$,
the proportion between the two being set by the value of $\dGam$.
\begin{footnote}
{
In the absence of noise, there may be various banding patterns that constitute
attainable steady states, corresponding to only imperceptibly different 
macroscopic stresses.
In this banding domain, the transient that lead to one of this or these steady-state solutions
essentially consists of two stages. First the fluidity evolves to form domains or bands 
of frozen ($a\sim 0$) and fluid ($a\sim a(\sigma^*)$) material, then in a second much slower stage these 
bands coarsen until a steady-state is reached.
}
\end{footnote}

Actually the physics is sometimes even richer. For some parameters,
usually close to the frozen branch - plateau region (left shaded area in
Figure \ref{fig11}b), an oscillating situation is reached
for both $a(z,t)$ and $\Sigma$: there is thus no steady-state ! 
Looking at the structure of $a(z,t)$, these oscillations correspond to
a {\em localized} oscillation of $a$ in a narrow region, often but not always
close to a wall, that alternatively freezes, so that the stress increases beyond $\sigma_Y$ then 
fluidizes which induces a release of the stress that drops below $\sigma_m$ which induces in turn
the re-freezing. This localized oscillating fracture, results in a macroscopic 
stick-slip behaviour.

The overall behaviour is very similar to that observed by Pignon et al. \cite{pig1}
on a laponite suspension (an 
oscillating fracture was observed by direct imaging), and by Varnik et al. in their simulations
\cite{bar1}. It should however be noted that in the situation where macroscopic bands are present
these works show some evidence of a value of $\dgam$ in the flowing band that changes with
$\dGam$ (and thus the size of this flowing region), in contrast to the predictions
of the present simple model.

\subsubsection{Transients}

Of course there are in principle lots of thing that can be analyzed beyond
the long-time response to an applied constant stress or shear-rate.
Of particular interest could be the analysis of the mechanical aging of a system prepared in
a heterogeneous banded state. 
\begin{footnote}
{
For a mechanical quench to study mechanical aging (see 2.2), 
one should clearly shear 
the system strongly so that it lies fully on the fluidized branch.
}
\end{footnote}

\subsection{Intermediary conclusion}

From experimental observations
emerges the picture that for some colloidal soft solids,
the structure of slow flows can be complex, both temporally and spatially.
To apprehend this complexity various experimental schemes,
possibly going beyond the measurement of purely mechanical
responses may be necessary. Eventually, experimental, analytic and
numerical works emphasize the role of the boundaries,
always present in a mechanical measurement .

\section{Criticism of the model and perspectives}

We end with a brief description of the interests and shortcomings of the
models described by equations (\ref{e1}) and (\ref{e2}),
before presenting a few comments as to some perspectives.

\subsection{A classification of the phenomenologies}

We have built a very simple class of models, where the main 
ingredients of the physics at work is hidden
in the functions $f$ and $g$.
Analyzing these models a classification has naturally emerged
related to the behaviour of these functions at low fluidity.

\begin{itemize}

\item
A choice for $f$ determines the value of $\alpha$ and thus of
of $\mu=(\alpha-1)^{-1}$.

- If $\alpha >2$ the system undergoes spontaneously sub-aging ($\mu<1$),
with the corresponding rescaling of the stress relaxation modulus, in
qualitative accordance with some experiments \cite{der2,ram2}.

- On the contrary $\alpha <2$ imposes super-aging $\mu>1$,
and stresses that freeze without relaxing to $0$. The limit case $\alpha=1$ corresponds 
to an exponential temporal increase of the mechanical relaxation time, as observed
in \cite{bon3}. 

- The marginal full-aging case $\alpha=2, \mu=1$ is consistent with other
observations \cite{clo1,via2}.

\item
The combined choice of $f$ and $g$ sets the value of
$\epsilon$. 
\begin{footnote}
{
Once again, a realistic model according to (\ref{e1}-\ref{e4})
forbids $\epsilon <0$, as this would mean resistance of the frozen
state ($a=0$,$\dgam=0$) to an infinite amount of stress.
}
\end{footnote}

- If $\epsilon >0$, the steady-state flow curve  
is that of a ``gentle'' power-law fluid, that flows homogeneously, and
can be confused with a yield-stress fluid only due to experimental difficulties
to access the weak $\dgam$ domain.

- If $\epsilon =0$, the system has a true yield stress,
and a Herschel-Buckley behaviour $\sigma
( \dot{\gamma}\rightarrow 0)\simeq \sigma_Y + A \dot{\gamma}^{n}$.
If $A \ge 0$ the system again is expected to flow homogeneously,
at a steady rate $\dgam$ if the stress is larger than $\sigma_Y$. 
However, $A$ can be negative,
in which case there is an initial decreasing branch in the
flow curve, yielding the viscosity bifurcation, gradient banding
and sometimes oscillating phenomenologies described in section 4.
\begin{footnote}
{
There is then another important characteristic state
($a_m, \dgam_m,\sigma_m$) (see Figure 10) that may or not be in the range 
of small fluidities that allow the expansions of (\ref{e3}-\ref{e4}).
}\end{footnote}

\end{itemize}
The classifications according to the signs of $\alpha-2$ and $\epsilon$ are a priori 
not directly related to one another which allows in principle for the existence of various
kind of soft-solids from a phenomenological point of view. 
In both the scalar model of subsection 3.2.1
and in the activated model of Lemaitre (3.2.2), the exponents $\alpha$ and $\epsilon$ are related by
$\alpha - 2=\epsilon$,
so that the two classifications above boil down to a single one.
In the 4-state model presented in 3.2.3, $\alpha=1$ and $\epsilon=0$ corresponding to
a super-aging yield-stress fluid.
The SGR model \cite{sol1}, 
which cannot be mapped onto the simple equations (\ref{e1},\ref{e2}),
yields a phenomenology corresponding
to full aging ($\alpha=2$) and a finite yield stress ($\epsilon=0$).

This diversity of possible behaviour seems indeed to show up in the spectrum of ``soft solids''.
Very crudely what people often name ``gels'' 
typically fall in the super-aging ($\alpha <2$)
and yield-stress behaviour ($\epsilon=0$) category, whereas ``glasses'' often 
display sub-aging or aging behaviour, with a rheology often difficult to discriminate
between a power-law behaviour with a weak exponent or a true yield stress-behaviour. 
I recall however (see section 1) that there is no clear definition of these terms,
and that tuning of some parameters (ionic strength, temperature, etc ..)
can lead continuously from one category to the other.

\subsection{Successes and failures of these models}

An potential interest of the models encompassed 
by equations (\ref{e1},\ref{e2})  is their analytical simplicity which
could allow to use them to follow or anticipate complex
rheological histories.

A more thorough  attempt to confront experimental data
with a specific set of exponents in equations 
 (\ref{e3},\ref{e4}) was performed for suspensions of protected
silica particles \cite{der4}, with the following
set of successes and failures:

Successes: the equations naturally describe mechanical aging
with a rescaling variable 
$S(t',t_w)=\frac{(t'+t_w)^{1-\mu}-t_w^{1-\mu}}{(1-\mu)\tau_0^{1-\mu}}$
naturally showing  up in equation (\ref{eq6}), that 
is well suited to account for the rescaling of relaxation experiments performed 
 at different waiting times on rather diverse systems \cite{clo1,der2,ram2}.
Also many non-linear features of steady and oscillatory rheology are reasonably captured
with rather simple forms for $f$ and $g$. A nice point is that the model also apprehends 
semi-quantitatively correctly features of transient regimes \cite{der4}, and
as such can be of help to anticipate/interpret experiments.

Failures: the use of a single (although evolving) time scale leads to inadequate
scaling functions for the mechanical response. For example equation 
 the modulus in (\ref{eq6}) is a single exponential of the rescaling variable,
whereas experiments in \cite{der2,ram2} suggest a stretched exponential (in $S(t',t_w)$)
(see e.g. fig. \ref{figexp}).
In addition, in many systems \cite{bor2,der4} the value of the elastic plateau, which is 
a constant $G_0$ in the model,
actually increases slowly with time as $\log(t_w)$. This usually mild feature can in principle be accounted
for by a natural modification of the system of equations
if one thinks in terms of activation barriers ($G_0=G_0({\cal E})$ with $\cal E$ increasing as $\log(t_w)$
in the description
of 3.2.2 , see also \cite{via2}), but at the cost of an increased complexity.
Eventually the description of the vicinity of the fluid/soft-solid transition
using only $r_1$ as a control parameter was shown to be unsatisfactory again
because it neglects the variations of $G_0$ with concentration.

So altogether, the main merit of these models is indeed that their simplicity and tractability
permits simple estimates for the outcome of an experiment.
They also allow one to make qualitative comments on the influence
of heterogeneities in banding flow, in a domain where more accurate/complete models
may be practically unusable.

A clear drawback at this stage is the total absence of connection of the phenomenological
equations and exponents with microscopic mechanisms of some generality,
or with coarse-grained versions of the latter. It is at this level that a possible relation
between the microscopic interactions and the macroscopic rheological behaviour
is to be made.

In the following we comment briefly on direction for possible extensions of these models,
very likely at the cost of simplicity.

\subsection{Better models: more variables? which collective physics ?}

Taking the equations  (\ref{e1},\ref{e2}) as a starting point,
extensions of the number and type of variables
are suggested by experimental observations and theoretical considerations:

-  The observation of stretched exponentials response functions \cite{der2,ram2}
(in the rescaled variables) suggests the existence of a spectrum of relaxation time
rather than a single one. For homogeneous systems
this suggests the use of instantaneous 
distributions or probability distributions
for either the local deformation (as e.g. in the SGR) or the local stress (as e.g. in \cite{heb1,der1,deb1}),
rather than the local averages $\sigma$ and $\dgam$ in the models quoted here. 

- Recent experimental observations have shown that an applied moderate 
shear can in some special conditions lead to an actual increase of the apparent mechanical 
age of the system \cite{via1,via2}. To account for such a flow induced 
``overaging'' it is tempting to evoke a set
of processes operating in parallel, in line with the previous point.

- In the observations of Cloitre et al. \cite{clo1}, a logarithmic recovery of the strain
is observed when stress is cut down to $0$. Such an evolution is forbidden
in the Maxwell model which implicitly takes $\dgam(t)$ as the driving and $\sigma(t)$
as the response. A way out if one insists on avoiding using whole distributions
is to introduce at least a single ``tensorial'' variable in addition to the stress,
which keeps the memory of the direction of past flow (a task which the ``scalar'' fluidity
$a$ is unable to perform). I point out that in related models for the plastic flow of amorphous
solids, recent developments by Langer and co-workers actually suggest
the use of such local variables to describe the instantaneous state of the system
(``shear-transformation-zones'' models \cite{fal1,lan1,eas1,lem1}).
Of course models involving local distributions of stresses or strains will naturally 
bear such ``tensorial'' memory.

- To take into account the banding observed in some cases and the corresponding
localization of the deformation, it is necessary to extend whatever is
the set of local variables to spatially and temporally dependent fields.
Of course this can be performed in the simple fluidity model presented here 
\cite{pic1}, but should in principle be done for models involving local distributions.

- An eventual point in this list is that the description we have made up to know
has neglected the intrinsically tensorial nature of both stress, deformation, and structure or texture
factors. There are instances where normal stresses have been shown to play a role in the selection
of heterogeneous patterns \cite{sch1}, and situations where the convection of fluctuations 
or structures by the flow is important require a proper tensorial description 
(see e.g. the recent mode-coupling description of \cite{fuc1}).

As a relevant aside, let me emphasize that
the respective role of stress and strain (or strain rate)
in constitutive mechanical models is far from clear. 
Most models actually take one or the other as the driving
field, the remaining one acting as a state variable. In an experiment,
what is actually imposed is in general boundary conditions on one or the other,
with the propagation of these 
tensorial fields in the bulk ruled by mechanical equilibrium and conservation laws,
in addition to the constitutive rule chosen.

At the end of this list one expects thus that a general model should consist
in a set of coupled non-linear differential equations (or possibly worse
integro-differential),  describing a large number of
spatially and temporally varying fields, with consequently very diverse possible 
outcomes even under steady driving \cite{gro1,gro2,cat1}.
Such models could remain rather phenomenological (as e.g. the SGR model)
or try to make tighter connections with structural information 
on the colloidal system (as e.g. in models inspired by mode-coupling theories \cite{fuc1}).
At a very local scale, one expects very dramatic non-linear
events, the ``rearrangements'', to be the building elements of a more collective
macroscopic response.

It is also possible that for some problems involving heterogeneous flows,
it could be physically more meaningful to describe
the evolution by focusing on a dynamical
description of the interfaces between the ``dynamical phases'' in contact,
rather than carrying a complete description
of the fields everywhere in the bulk
\cite{olm1,ajd1,gov1,but1}.

\subsection{Back to facts}

Given the complexity alluded to above in the construction of models,
it seems that the most important thing to do now is,
guided by concepts and simple models that have emerged in the last years,
to build first sets of experimental data on well controlled systems,
where one has grasp on the synthesis of the system
and experimental means of investigation of their structure under flow
beyond the simple mechanical response (see e.g. the review \cite{cip1}).
A similar statement holds for numerical simulations \cite{bar1,deb1,liu2}.
From the discussion above, one should obviously not necessarily expect
too much universality but rather try to infer classes of behaviour
from the analysis of such sets of data.

\acknowledgements
I am very grateful to D. Bartolo, L. Bocquet, G. Picard and V. Viasnoff
for their critical comments on earlier versions of these notes.



 \end{document}